\documentclass[12pt]{article}

\usepackage{amsfonts, graphicx,caption,subfig}

\usepackage{amssymb}
\usepackage{amsmath}
\usepackage{amsfonts}

\usepackage{epsfig}

\def\be{\begin{equation}}
\def\ee{\end{equation}}
\def\ben{\begin{displaymath}}
\def\een{\end{displaymath}}
\def\ba{\begin{array}{c}}
\def\ea{\end{array}}

\newcommand{\bea}{\begin{eqnarray}}
\newcommand{\eea}{\end{eqnarray}}

\newcommand{\kt}{\rangle}
\newcommand{\br}{\langle}
\newcommand{\ed}{\end{document}}

\begin{document}

\titlepage

  \begin{center}{\Large \bf

New Concept of Solvability in Quantum Mechanics

 }\end{center}

\vspace{5mm}

  \begin{center}

Miloslav Znojil\footnote{ e-mail: znojil@ujf.cas.cz}

 \vspace{3mm}

Nuclear Physics Institute ASCR, 250 68 \v{R}e\v{z}, Czech
Republic\\

 \vspace{5mm}

\end{center}

\vspace{5mm}
\section*{Abstract}

In a pre-selected Hilbert space ${\cal H}$ of quantum states
$|\psi\kt \in {\cal H}$ the unitarity of the evolution is usually
guaranteed via a pre-selection of the generator (i.e., of the
Hamiltonian operator $H$) in self-adjoint form, $H= H^\dagger$. In
fact, the simultaneous use of both of these pre-selections is
overrestrictive. One should be allowed to make a given Hamiltonian
self-adjoint only after an {\em ad hoc} generalization of Hermitian
conjugation, $H^\dagger \to H^\ddagger :=\Theta^{-1}H^\dagger
\Theta$. We argue that in the generalized, hidden-Hermiticity
scenario with nontrivial metric $\Theta\neq I$, the current concept
of solvability (meaning, most often, the feasibility of a
non-numerical diagonalization of $H$) requires a generalization
allowing for a non-numerical form of $\Theta$. A few illustrative
solvable quantum models of this type are presented.

 \vspace{9mm}

\noindent PACS  03.65.Db, 03.65.Aa, 05.30.Rt, 02.30.Sa, 03.65.Ca,
11.30.Er, 21.60.Ev, 31.15.xt



\vspace{9mm}

  \begin{center}
\end{center}


\section{Introduction}

In our recent paper \cite{fortschr} it has been noticed that
contrary to the current belief, an active use of an {\em ad hoc}
variability of the inner products (i.e., in other words, of the
freedom of choosing a nontrivial metric $\Theta \neq I$ in the
correct physical Hilbert space of quantum states ${\cal H}^{(S)}$
where the superscript $^{(S)}$ stands for ``standard'') is {\em not}
restricted to the so called non-Hermitian quantum mechanics and to
its characteristic applications in nuclear physics \cite{Geyer} or
in molecular physics \cite{Bishop} or in the relativistic quantum
kinematical regime \cite{kg} or in the ${\cal PT}-$symmetric quantum
dynamical regime \cite{Carl}. In our present paper we intend to
develop this idea and to describe some of its consequences in some
detail.

%

From the point of view of the recent history of quantum mechanics it
was, certainly, fortunate that in some of the above-mentioned
specific hidden-Hermiticity contexts people discovered the
advantages of working with such an operator representation $H$ of a
given observable quantity (say, of the energy) which only proved
Hermitian {\em after} a change of the inner product in the initially
ill-chosen (i.e., by assumption, unphysical) Hilbert space ${\cal
H}^{(F)}$ (the superscript $^{(F)}$ might be read here as
abbreviating ``former'', ``first'',  ``friendly'' or, equally well,
``false'' \cite{SIGMA}). Let us emphasize that the shared motivation
of many of the above-cited papers speaking about non-Hermitian
quantum mechanics resulted just from the observation that several
phenomenologically  interesting operators (say, Hamiltonians) $H$
appear manifestly non-Hermitian in the ``usual'' textbook setting
and that they only become Hermitian in some much less common
representation of the Hilbert space of states.

The amendments of space were, naturally, mediated by the mere
introduction of a non-trivial metric $\Theta=\Theta^{(S)}\neq I$
entering the upgraded, $S-$superscripted inner products,
 \be
 \br \psi_1|\psi_2\kt^{(F)}\
 \ \to \
 \br \psi_1|\psi_2\kt^{(S)}\
 :=\br \psi_1|\Theta|\psi_2\kt^{(F)}
 \,.
 \label{ip}
 \ee
Such an inner-product modification changed, strictly speaking, the
Hilbert space, ${\cal H}^{(F)}\to {\cal H}^{(S)}$. This had {several
independent} reasons. Besides the formal necessity of re-installing
the unitarity of the evolution law, the costs of the transition to
the more complicated metric were  found more than compensated by the
gains due to the persuasive simplicity of Hamiltonian (cf.
\cite{Geyer} or \cite{Carl} in this respect). Moreover, for some
quantum systems the transition $ F \to S$ may prove motivated by
physics itself. The most elementary illustration of such a
fundamental reason can be found in our recent study \cite{confser}
where a consistent simulation of the cosmological phenomenon of
quantum Big Bang has been described. In the model the Hamiltonian
remained self-adjoint in the false space, $H = H^\dagger$. Still,
another relevant observable proved non-Hermitian there, $Q \neq
Q^\dagger$.


%
\begin{figure}
\includegraphics[width=8cm,angle=-90]{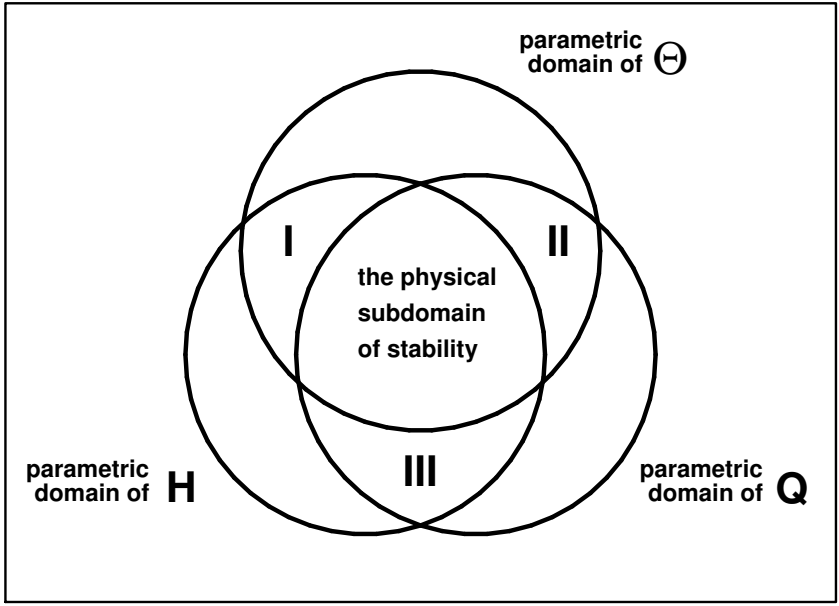}
\caption{Generic domains of parameters for which the metric $\Theta$
exists (upper disc) or for which the spectra of observables $H$ or
$Q$ remain potentially observable (the two respective lower discs).}
\label{obr0a}
\end{figure}


%


%

For an entirely general quantum system characterized by two
observables $H$ and $Q$, Hermitian or not, a fully universal
scenario may be found displayed in Fig.~\ref{obr0a}. In the picture
(where the whole plane symbolizes a multidimensional space of all
parameters of the model) we see three circles. Schematically, they
represent three boundaries $\partial {\cal D}$ of  three domains $
{\cal D}$. Thus, the spectrum of $H$ is assumed potentially
observable (i.e., real and non-degenerate) in the left lower domain
$ {\cal D}_H$. Similarly, the spectrum of $Q$ is real and
non-degenerate inside the right lower domain $ {\cal D}_Q$. In
parallel, the spectrum of the available Hermitizing metrics $\Theta$
must be, by definition, strictly positive (upper circle, domain  $
{\cal D}_\Theta$). In this arrangement, operator $Q$ ceases to
represent an observable in domain ``I'' while operator $H$ ceases to
represent an observable in domain ``II''.  In domain ``III'',
neither of these two operators can be made Hermitian using the
available class of metrics $\Theta$, in spite of the reality of both
spectra.

A number of open questions emerges. Some of them will be discussed
in our present paper. Via a few illustrative examples we shall show,
among other, that and why the variability of the metric $\Theta$ in
the physical Hilbert space ${\cal H}^{(S)}$ represents an important
merit of quantum theory and that and why the closed-form
availability of operator $\Theta$ (i.e., a new form of solvability)
is of a truly crucial importance in applications.

\section{Methodical guidance: dimension two\label{toymo}}

\subsection{Toy-model Hamiltonian}

In the simplest possible two-dimensional and real Hilbert space
${\cal H}^{(F)}\ \equiv\ \mathbb{R}^2$ an instructive sample of the
time evolution may be chosen as generated by the Hamiltonian (i.e.,
quantum energy operator or matrix) of Ref.~\cite{maximal},
 \be
 H=H^{(2)}(\lambda)=
 \left [
 \begin{array}{cc}
 -1&\lambda\\
 -\lambda&1
 \ea
 \right ]\,.
 \label{henry}
 \ee
Its eigenvalues $E_\pm^{(2)} = \pm \sqrt{1-\lambda^2}$ are
non-degenerate and real (i.e., in principle, observable) for
$\lambda$ inside interval $(-1,1)$. On the two-point domain boundary
$\{-1,1\}$, these energies degenerate. Subsequently, they complexify
whenever  $|\lambda|>1$. In the current literature one calls the
boundary points $\lambda = \pm 1$ ``exceptional points'' (EP,
\cite{Kato}). At these points the eigenvalues degenerate and our
toy-model Hamiltonian ceases to be diagonalizable, becoming
unitarily equivalent to a triangular Jordan-block matrix,
 \ben
 H^{(2)}(1)=
 \left [H^{(2)}(-1)\right ]^\dagger=
 \left [
 \begin{array}{cc}
 -1&1\\
 -1&1
 \ea
 \right ]
 =
 \een
 \ben
 =\frac{1}{2}\,
 \left [
 \begin{array}{cr}
 1&-1\\
 1&1
 \ea
 \right ]\,
 \left [
 \begin{array}{cc}
 0&1\\
 0&0
 \ea
 \right ]\,
 \left [
 \begin{array}{rc}
 1&1\\
 -1&1
 \ea
 \right ]
 \,.
 \label{ouhenry}
 \een
At $|\lambda|>1$, the diagonalizability gets restored but the
eigenvalues cease to be real, $E_\pm^{(2)} = \pm {\rm
i}\,\sqrt{\lambda^2-1}$. In the spirit of current textbooks, this
leaves these purely imaginary complex conjugate energies
unobservable.

\subsection{Hidden Hermiticity: The set of all eligible metrics}

Our matrix $H^{(2)}(\lambda)$ remains diagonalizable and
crypto-Hermitian whenever $-1<\lambda=\sin \alpha<1$, i.e., for the
auxiliary Hamiltonian-determining parameter $\alpha$ lying inside a
well-defined physical domain ${\cal D}_H$ such that $\alpha \in
(-\pi/2,\pi/2)$. In such a setting, matrix $H^{(2)}(\lambda)$
becomes tractable as a Hamiltonian of a hypothetical quantum system
whenever it satisfies the above-mentioned hidden Hermiticity
condition
 \be
 H= H^\ddagger:= \Theta^{-1}H^\dagger\Theta\,.
\
 \label{hiherua}
 \ee
The suitable candidates for the Hilbert-space metric are all easily
found from the latter linear equation,
 \be
 \Theta=\Theta^{(2)}_\lambda(a,d)=
 \left [
 \begin{array}{cc}
 a&b\\
 b&d
 \ea
 \right ]
 \,, \ \ \ \ \
 b=-\frac{\lambda}{2}(a+d)
 \,.
 \label{merie}
 \ee
All of their eigenvalues must be real and positive,
 \be
 \theta_\pm = \frac{1}{2}
 \left [ a+d\pm \sqrt{(a-d)^2+\lambda^2(a+d)^2}
 \right ]>0\,.
 \label{eigvel}
 \ee
This is satisfied for any positive $\sigma=a+d>0$ and with any real
$\delta=a-d$ such that
 \be
 \sqrt{1-\lambda^2}= \cos \alpha \ \ > \ \
  \frac{\delta}{\sigma} \ \ > \ \, -\sqrt{1-\lambda^2}= - \cos
 \alpha\,.
 \ee
Without loss of generality we may set $\sigma=2$, put $\delta= \cos
\alpha \cos \beta$ and treat the second free parameter $\beta \in
(-\pi/2,\pi/2)$ as numbering the admissible metrics
 \be
 \Theta^{(2)}_{(physical)}=
 \left [
 \begin{array}{cc}
 1+\cos \alpha \cos
\beta &-\sin \alpha \\
 -\sin \alpha&1-\cos \alpha \cos
\beta
 \ea
 \right ]\,
 \ee
with eigenvalues
 \be
 \theta_\pm = 1\pm \sqrt{1- \cos^2 \alpha \sin^2
\beta}
 >0\,.
 \ee
Thus, all of the eligible physical Hilbert spaces are numbered by
two parameters, ${\cal H}^{(S)}={\cal H}^{(S)}(\alpha,\beta)$.


%

\subsection{The second observable $Q = Q^\ddagger$}

What we now need is the specification of the domain ${\cal D}_Q$.
For the general four-parametric real-matrix ansatz
 \be
 \tilde{{Q}}=
 \left [
 \begin{array}{cc}
 w&x\\
 y&z
 \ea
 \right ]
  \,
 \ee
the assumption of observability implies that the eigenvalues must be
both real and non-degenerate,
 \be
 4xy > -(w-z)^2\,.
 \ee
Once we shift the origin  and rescale the units we may set, without
loss of generality, $w=-z=-1$. This simplifies the latter condition
yielding our final untilded two-parametric ansatz
 \be
 {Q}=
 \left [
 \begin{array}{cc}
 -1&x\\
 y&1
 \ea
 \right ]
  \,,\ \ \ \ xy>-1\,.
  \label{equuu}
 \ee
At any fixed metric $\Theta^{(2)}_{(physical)}$ the
crypto-Hermiticity constraint (\ref{hiherua}) imposed upon matrix
(\ref{equuu}) degenerates to the single relation
 \be
  x-y=2 \sin \alpha-(x+y)\cos \alpha \cos \beta \,.
  \label{rela}
 \ee
The sum $s=x+y$ may be now treated as the single free real variable
which numbers the eligible second observables. The range of this
variable should comply with the inequality in Eq.~(\ref{equuu}).
After some straightforward additional calculations one proves that
the physical values of our last free parameter remain unrestricted,
$s \in  \mathbb{R}$, due to the validity of Eq.~(\ref{rela}). We may
conclude that our example is fully non-numerical. It also offers the
simplest nontrivial explicit illustration of the generic pattern as
displayed in Fig.~\ref{obr0a}.


\section{Hilbert spaces ${\cal H}^{(F)}$ of dimension $N$\label{toymoN}}

\subsection{Anharmonic Hamiltonians}

During the developments of mathematics for quantum theory, one of
the most natural paths of research started from the exactly solvable
harmonic-oscillator potential $V^{(HO)}(x)=\omega^2x^2$ and from its
power-law perturbations $V^{(AHO)}(x)=\omega^2 x^2+g\,x^m$.
Perturbation expansions of energies proved available even at the
``unusual'', complex values  of the coupling constants $g \notin
\mathbb{R}^+$. The particularly interesting mathematical results
have been obtained at $m=3$ and at $m=4$. In physics and, in
particular, in quantum field theory the climax of the story came
with the letter \cite{BB} where, under suitable {\em ad hoc}
boundary conditions and constraints upon $g=g(m)$ (called,
conveniently, ${\cal PT}-$symmetry), the robust reality (i.e., in
principle, observability) of the spectrum has been achieved at any
real exponent $m>2$ even for certain unusual, complex values of the
coupling.

It has been long believed that the ${\cal PT}-$symmetric
Hamiltonians $H=H(m)$ with real spectra are all consistent with the
postulates of quantum theory, i.e., that these operators are
crypto-Hermitian, i.e., Hermitian in the respective
Hamiltonian-adapted Hilbert spaces ${\cal H}^{(S)}(m)$ \cite{Carl}.
Due to the ill-behaved nature of the wave functions at high
excitations, unfortunately, such a simple-minded physical
interpretation of these models has been shown
contradictory~\cite{Siegl}. On these grounds one has to develop some
more robust approaches to the theory for similar models in the
nearest future.

In our present paper we shall avoid such a danger by recalling the
original philosophy of Scholtz et al~\cite{Geyer}. They simplified
the mathematics by admitting, from the very beginning, that just the
bounded-operator and/or discrete forms of the eligible
anharmonic-type toy-model Hamiltonians $H \neq H^\dagger$ should be
considered.

\subsection{Discrete  Hamiltonians}

For our present illustrative purposes we intend to recall, first of
all, one of the most elementary versions of certain general,
$N-$dimensional matrix analogues of the differential toy-model
Hamiltonians, which were proposed in Refs.~\cite{maximal}. Referring
to the details as described in that paper, let us merely recollect
that these Hamiltonians are defined as certain tridiagonal and real
matrices $H^{(N)}=H_0^{(N)}+V^{(N)}$ where the ``unperturbed'',
harmonic-oscillator-simulating main diagonal remains equidistant,
$H^{(N)}_{11}=(H_0^{(N)})_{11}=-N+1$,
$H^{(N)}_{22}=(H_0^{(N)})_{22}=-N+3$, \ldots,
$H^{(N)}_{NN}=(H_0^{(N)})_{NN}=N-1$ while the off-diagonal
``perturbation''  becomes variable and, say, antisymmetric,
$V^{(N)}_{12}=-V^{(N)}_{21}$, $V^{(N)}_{23}=-V^{(N)}_{32}$, \ldots
$V^{(N)}_{N-1N}=-V^{(N)}_{NN-1}$. The word ``perturbation'' is
written here in quotation marks because, in the light of results of
Ref.~\cite{tridiagonal}, the spectral properties of the model become
most interesting in the strongly non-perturbative regime where one
up-down symmetrizes and re-parametrizes the perturbation
$V^{(N)}_{k,k+1}=-V^{(N)}_{k+1,k}=$
 \ben
 =\sqrt{k(N-k)(1-t-t^2-\ldots-t^{J-1}-G_kt^J)}\,,
 \label{parameo}
 \een
 \ben
 \ \ \ \ \ \  \
 \ \ \ \ \ \  \
  N = 2J \ {\rm or }\ N=2J+1\,.
 \een
This parametrization proved fortunate in the sense that it enabled
us to replace the usual numerical analysis by a rigorous
computer-assisted algebra. In this sense, the model in question
appeared to represent a sort of an exactly solvable model, precisely
in the spirit of our present message.

The new parameter $t\geq 0$ is auxiliary and redundant. It may be
interpreted, say, as a measure of distance of the system from the
boundary $\partial {\cal D}_H$ of the domain of spectral reality. At
very small $t$ the local part of boundary $\partial {\cal D}_H$ has
been shown to have the most elementary form of two parallel
hyperplanes in the $J-$dimensional space of parameters
$G_n$~\cite{tridiagonal}.

In the simplest nontrivial special case of $N=2$ the present
Hamiltonian $H^{(N)}$ degenerates precisely to the above-selected
toy-model of section \ref{toymo}. {\em Vice versa}, the basic
components of the $N=2$ discussion (i.e., first of all, the
feasibility of the construction of the metric and of the second
observable) might be immediately transferred to all $N>2$. Several
steps in this direction may be found performed in our recent paper
on the solvable benchmark simulations of the phase transitions
interpreted as a spontaneous ${\cal PT}-$symmetry
breakdown~\cite{catast}.

\section{The problem of non-uniqueness of the {\em ad hoc}
metric $\Theta=\Theta({H})$}


The roots of the growth of popularity of the description of {\em
stable} quantum systems using representations of observables which
are non-Hermitian in an auxiliary Hilbert space ${\cal H}^{(F)}$ may
be traced back not only to the entirely abstract mathematical
analyses of spectra of quasi-Hermitian operators \cite{Dieudonne}
and of the operators which are self-adjoint in the so called Krein
spaces with indefinite metric \cite{Langer} but also to the
emergence of manageable non-Hermitian models in quantum field theory
\cite{BM} or even in classical optics \cite{Makris}, etc.

After a restriction of attention to quantum theory, the key problem
emerges in connection with the ambiguity of the assignment $H \to
\Theta(H)$ of the physical Hilbert space ${\cal H}^{(S)}$ to a given
generator $H$ of time evolution. For many phenomenologically
relevant Hamiltonians $H$ it appeared almost prohibitively difficult
to define and construct at least some of the eligible metrics
$\Theta=\Theta(H)$ in an at least approximate form (cf., e.g., Ref.
\cite{cubic} in this respect). Clearly, in methodical analyses the
opportunity becomes wide open to finite-dimensional and solvable toy
models.

\subsection{Solvable quantum models with more than one observable}

Let us restrict the scope of this paper to the quantum systems which
are described by a Hamiltonian $H=H(\lambda)$ accompanied by a
single other operator $Q=Q(\varrho)$ representing a complementary
measurable quantity like, e.g., angular momentum or coordinate. In
general we shall assume that symbols $\lambda$ and $\varrho$
represent multiplets of coupling strengths or of any other
parameters with an immediate phenomenological or purely mathematical
significance. We shall also solely work here with the
finite-dimensional matrix versions of our operators of observables.

In such a framework it becomes much less difficult to analyze one of
the most characteristic generic features of crypto-Hermitian models
which lies in their ``fragility'', i.e., in their stability up to
the point of a sudden collapse. Mathematically, we saw that the
change of the stability/instability status of the model is
attributed to the presence of the exceptional-point horizons in the
parametric space. In the context of phenomenology, people often
speak about the phenomenon of quantum phase transition
\cite{Makris}.

Let us now return to Fig.~\ref{obr0a} where the set of the
phase-transition points pertaining to the Hamiltonian $H$ is
depicted as a schematic circular boundary $\partial {\cal D}_H$ of
the left lower domain inside which the spectrum of $H$ is assumed,
for the sake of simplicity, non-degenerate and completely real.
Similarly, the right lower disc or domain  ${\cal D}_Q$ is assigned
to the second observable $Q$. Finally, the upper, third circular
domain ${\cal D}_\Theta$ characterizes the parametric subdomain of
the existence of a suitable general or, if asked for, special class
of the eligible candidates $\Theta$ for a physical metric operator.
The key message delivered by Fig.~\ref{obr0a} is that at any $N$,
the correct physics may still only be formulated inside the
subdomain ${\cal D}={\cal D}_H \bigcap {\cal D}_H \bigcap {\cal
D}_H$. A generalization of this scheme to systems with more
observables, $Q \to Q_1$, $Q_2$, \ldots \, would be straightforward.

\subsection{Quantum observability paradoxes }

One of the most exciting features of all of the above-mentioned
models may be seen in their ability of connecting the stable and
unstable dynamical regimes, within the same formal framework, as a
move out of the domain ${\cal D}$ though one of its boundaries. In
this sense, the exact solvability of the $N<\infty$ toy models
proves crucial since the knowledge of the boundary $\partial {\cal
D}_\Theta$ remains practically inaccessible in the majority of their
$N=\infty$ differential-operator alternatives \cite{cubic}.

In the current literature on the non-Hermitian representations of
observables, people most often discuss just the systems with a
single relevant observable $H(\lambda)$ treated, most often, as the
Hamiltonian. In such a next-to-trivial scenario it is sufficient to
require that  operator $H$ remains diagonalizable and that it
possesses a non-degenerate real spectrum. Once we add another
observable $Q$ into considerations, the latter conditions merely
specify the interior of the leftmost domain ${\cal D}_H$ of our
diagram Fig.~\ref{obr0a}.

One may immediately conclude that the physical predictions provided
by the Hamiltonian alone (and specifying the physical domain of
stability as an overlap between ${\cal D}_H$ and the remaining upper
disc or domain ${\cal D}_\Theta$) remain heavily non-unique in
general. According to Scholtz et al \cite{Geyer} it is virtually
obligatory to take into account at least one other physical
observable $Q=Q(\varrho)$, therefore.

In opposite direction, even the use of a single additional
observable $Q$ without any free parameters may prove sufficient for
an exhaustive elimination of all of the ambiguities in certain
models \cite{Carl}. One can conclude that the analysis of the
consequences of the presence of the single additional operator
$Q=Q(\varrho)$ deserves a careful attention. At the same time,
without the exact solvability of the models, some of their most
important merits (like, e.g., the reliable control and insight in
the processes of the phase transitions) might happen to be
inadvertently lost.

\section{Adding the degrees of freedom}

\subsection{Embedding: $N=2$ space inside $N=3$ space}

In the spirit of Ref. \cite{return} a return to observability may be
mediated by an enlargement of the Hilbert space. For example, a
weak-coupling immersion of our matrices $H^{(2)}(\lambda)$ in their
three by three extension
 \be
 H^{(3)}=
  \left[ \begin {array}{ccc} -1&1+z&0\\\noalign{\medskip}-1-z&1&g
\\
\noalign{\medskip}0&-g&3\end {array} \right]
 \label{rexhenry}
 \ee
may be interpreted as a consequence of the immersion of the smaller
Hilbert space (where one defined Hamiltonian (\ref{henry})) into a
bigger Hilbert space. Via the new Hamiltonian (\ref{rexhenry}), the
old Hamiltonian becomes weakly coupled to a new physical degree of
freedom by the interaction proportional to a small constant $g$.

%
%
%
%
%

In a way discussed in more detail in our older paper~\cite{horizon},
the boundary of the new physical domain ${\cal D}_{H^{(3)}}$
coincides with the zero line of the following polynomial $G(z,g)$ in
two variables,
 \ben
 60\,{g}^{2}{z}^{2}-6\,z{g}^{4}-12\,
 {g}^{2}{z}^{3}-{z}^{6}-162\,z+27\,{g}^{2}-18\,{g}^{4}-{g}^{6}-
 \label{zerol}
 \een
 \ben
 -153\,{z}^{2}-3\,{g}^{4}{z}^{2}-3\,{g}^{2}{z}^{4}-6\,{z}^{5}-30\,
 {z}^{4}-80\,{z}^{3}+144\,z{g}^{2}.
 \een
The shape of this line is shown in Fig.~\ref{obr1a}.
%


%
\begin{figure}
\includegraphics[width=8cm,angle=-90]{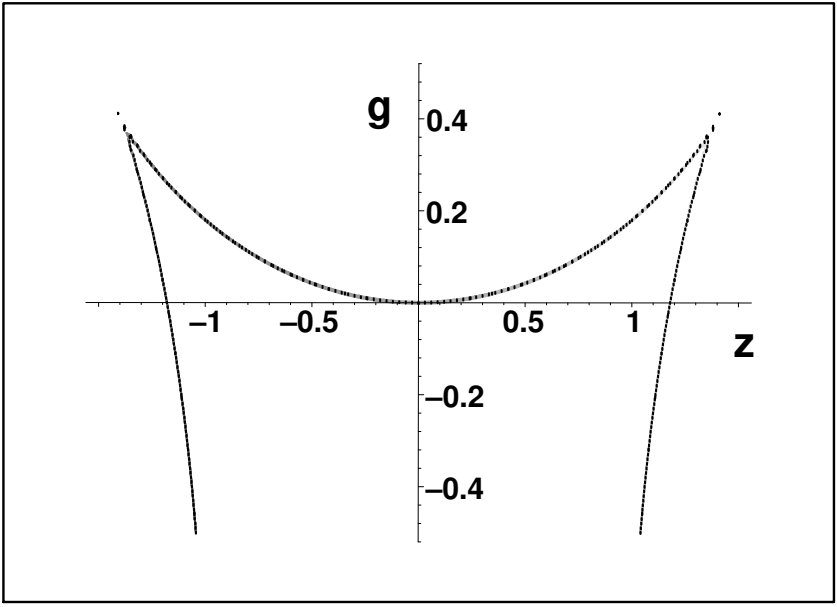}
\caption{The boundary of the domain of reality of the spectrum of
Hamiltonian (\ref{rexhenry}) in $z-g$ plane (i. e., the zero line of
polynomial $G(z,g)$).
 \label{obr1a}}
\end{figure}

In the vicinity of  $z=0$ and $g=0$, the truncated polynomial
$G_0(z,g)=27\,{g}^{2}-162\,z-18\,{g}^{4}
+144\,z{g}^{2}-{g}^{6}-153\,{z}^{2}-6\,z{g}^{4}$ appears useful as a
source of the auxiliary boundary of a fairly large subdomain ${\cal
D}_0$ of the physical domain ${\cal D}_{H^{(3)}}$ (cf.
Fig.~(\ref{obr2a})).



%
\begin{figure}
\includegraphics[width=8cm,angle=-90]{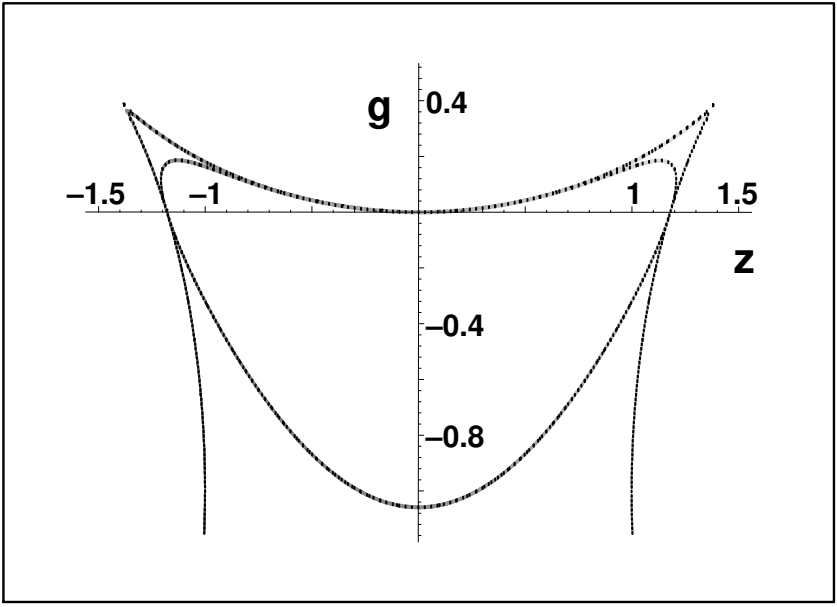}
\caption{The comparison of zero lines of functions $G_0(z,g)$ and
$G(z,g)$.
 \label{obr2a}}
\end{figure}

All of these observations imply that the original $z=0$ boundary
bends up, i.e., the net effect of the introduction of the new, not
too large coupling $g \neq 0$ lies in the enlargement of the domain
of the reality of the energy spectrum beyond $\lambda=1$ (and,
symmetrically, below $\lambda=-1$). In other words, an enhancement
of the stability of the system with respect to some random
perturbations is achieved simply by its coupling to an environment.

\subsection{Global metrics at $N=3$}

\begin{figure}
\includegraphics[width=8cm,angle=-90]{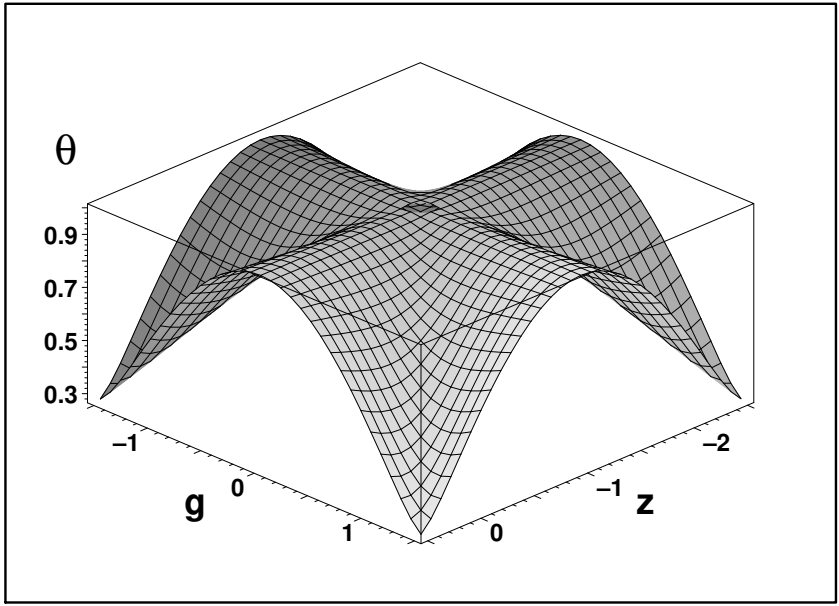}
\caption{The subdominant eigenvalue of metric $\Theta^{(3)}(1,1,1)$.
It stays safely positive in the whole preselected rectangle of
parameters $z$ and $s$.
 \label{obr3a}}
\end{figure}


%
\begin{figure}
\includegraphics[width=8cm,angle=-90]{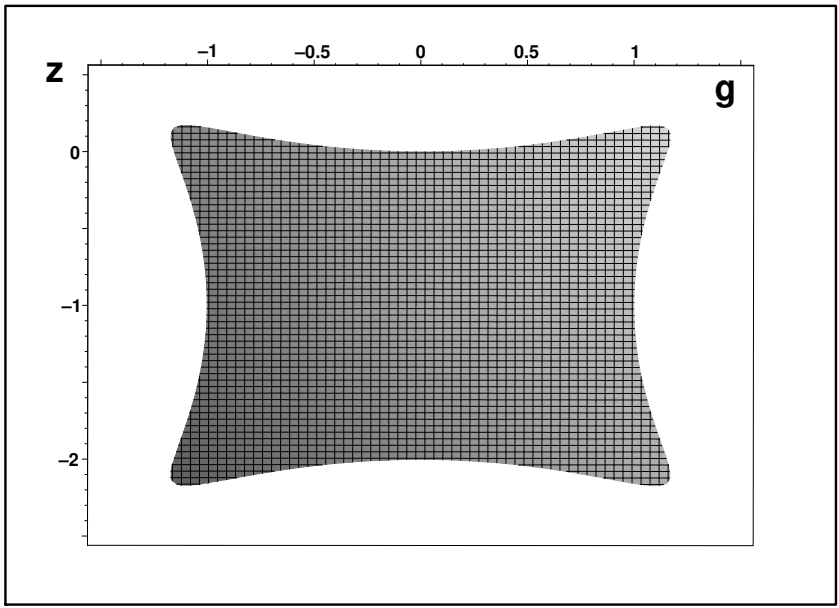}
\caption{The complete domain of positivity of the smallest
eigenvalue of metric $\Theta^{(3)}(1,1,1)$.
 \label{obr4a}}
\end{figure}

%

The enlarged system controlled by Hamiltonian $H^{(3)}$ of
Eq.~(\ref{rexhenry}) has been chosen as crypto-Hermitian. The
construction of the eligible metrics
 \be
 \Theta^{(3)}=
  \left[ \begin {array}{ccc} a&b&c\\\noalign{\medskip}b&f&h
 \\
 \noalign{\medskip}c&h&m\end {array} \right]
 \ee
of the enlarged and re-coupled system may be perceived as another
exercise in the construction of the metrics exactly, by
non-numerical means. Using the similar techniques we obtain,
step-by-step,
 $$
 c=(-h-h\,z-b\,g)/4,
 $$
 $$
 b=-(4\,a\,z+4\,f+4\,f\,z+4\,a+h\,g+g\,h\,z)/(8+g^2)
 $$
and eliminate, finally, $-2h\,(9+2\,z+z^2+g^2)/g=$
 $$
 =
 -2\,a\,z-a\,z^2+7\,f-2\,f\,z-f\,z^2-a+8\,m+m\,g^2+f\,g^2.
 $$
Thus, starting from the three arbitrary real parameters
$\Theta_{11}=a$, $\Theta_{22}=f$ and $\Theta_{33}=m$ we recursively
eliminate $\Theta_{1,3}=c=(-h-h\,z-b\,g)/4$,
$\Theta_{1,2}=b=-(4\,a\,z+4\,f+4\,f\,z+4\,a+h\,g+g\,h\,z)/(8+g^2)$
and $\Theta_{2,1}= h=
-1/2\,g\,(-2\,a\,z-a\,z^2+7\,f-2\,f\,z-f\,z^2-a+8\,m+m\,g^2
+f\,g^2)/(9+2\,z+z^2+g^2)$. As a final result we obtain the formula
for
 $$2\,(9+2\,z+{z}^{2}+{g}^{2} )\,\Theta_{1,2}=$$
 \ben
 =
 z{g}^{2}m+fz{g}^{2}+m{g}^{2}+f{g}^{2}-3\,f{z}^{2}-3\,a{z}^{2}-
 \een
 \ben
 -f{z}^{3}-a{z}^{3}-9\,a-11\,fz-11\,az-9\,f\,.
 \een
Thus, we may denote $\Theta=\Theta^{(3)}(a,f,m)$ and conclude that
the metric is obtainable in closed form so that our extended, $N=3$
quantum system remains also solvable.

If we also wish to determine the critical boundaries $\partial {\cal
D}_\Theta$ of the related metric-positivity domain ${\cal
D}_\Theta$, the available Cardano's closed formulae for the
corresponding three eigenvalues $\theta_j$ yield just the correct
answer in a practically useless form. Thus, we either have to recall
the available though still rather complicated algebraic
boundary-localization formulae of Ref.~\cite{horizon} or,
alternatively, we may simplify the discussion by the brute-force
numerical localization of a sufficiently large metric-supporting
subdomain in the parametric space. For the special choice of
$a=f=m=1$ we found, for example, that for the sufficiently large
range of parameters $z$ and $g$ as chosen in Figs.~\ref{obr3a} and
~\ref{obr4a} we reveal that while the two upper eigenvalues
$\theta_2$ and $\theta_1$ remain safely positive, the minimal
eigenvalue $\theta_0$ only remains positive inside the minimal
domain of positivity as displayed in Fig.~\ref{obr4a}. Thus, the
boundary of the latter domain represents an explicit concrete
realization of its abstract upper-circle representative in
Fig.~\ref{obr0a}.

\section{Up-down symmetrized couplings to the environment}

\subsection{Toy model with $N=9$}

The $\cal PT-$symmetric and tridiagonal nine-by-nine-matrix
Hamiltonian $ H^{(9)}$ of Ref.~\cite{maximal} reads
 $$
 \left[ \begin {array}{ccccccccc}
  -8&b& 0&0&0&0&0&0&0\\
 -b&-6&  c&0&0&0&0&0&0\\
  0& -c&   -4& d&0&0&0&0&0\\
  0&  0&  -d&-2& {\alpha}&0&0&0&0\\
  0&0&  0&  -{\alpha}&0& {\alpha}&0&0&0\\
  0&0&  0& 0& -{\alpha}&2& d&0&0\\
  0& 0& 0& 0& 0&-d&4& c&0\\
  0& 0& 0& 0& 0&0&-c&6&b\\
  0&0& 0& 0& 0& 0&0&-b&8\end
{array}
 \right]\,.
 $$
In the limit $\alpha\to 0$ it splits into a central one-dimensional
submatrix with eigenvalue $0$ and a pair of non-trivial four-by-four
sub-Hamiltonians $H^{(4)}$. The spectrum remains real, say, for the
family of parameters $b = \sqrt{3 + 3 t}$, $c = 2 \sqrt{1 + t}$ and
$d = \sqrt{3 + 3 t}$. They span an interval in the physical domain
${\cal D}_H$ whenever $t$ stays negative, $t \in (-\infty,0)$
\cite{tridiagonal}.

At $\alpha=0$ the special and easily seen feature of the latter
operator (i.e., matrix) is that at $t=0$ (i.e., at the boundary of
its physical domain ${\cal D}_H$) it ceases to represent an
observable because its eigenvalues degenerate. Indeed, the vanishing
level $E_4=0$ separates from the two degenerate quadruplets of
$E_{4+j}=- E_{4-j}=5$ with $j=1,2,3,4$. Subsequently, at $t > 0$,
these eigenvalues get, up to the constantly real level $E_4=0$,
complex. This makes the model suitable for quantitative studies of
the properties of the boundary $\partial {\cal D}_H$ \cite{catast}.

%
%

\subsection{Boundary $\partial {\cal D}_H$}

The $t-$independent level $E_4=0$ is a schematic substitute for a
generic environment. Each of the two remaining subsystems remains
coupled to this environment by the coupling or matrix element
${\alpha}$. We shall choose its value as proportional to $t$ via a
not too large real coupling constant $\beta$, ${\alpha}= \beta\,t$.


%
\begin{figure}
\includegraphics[width=8cm,angle=-90]{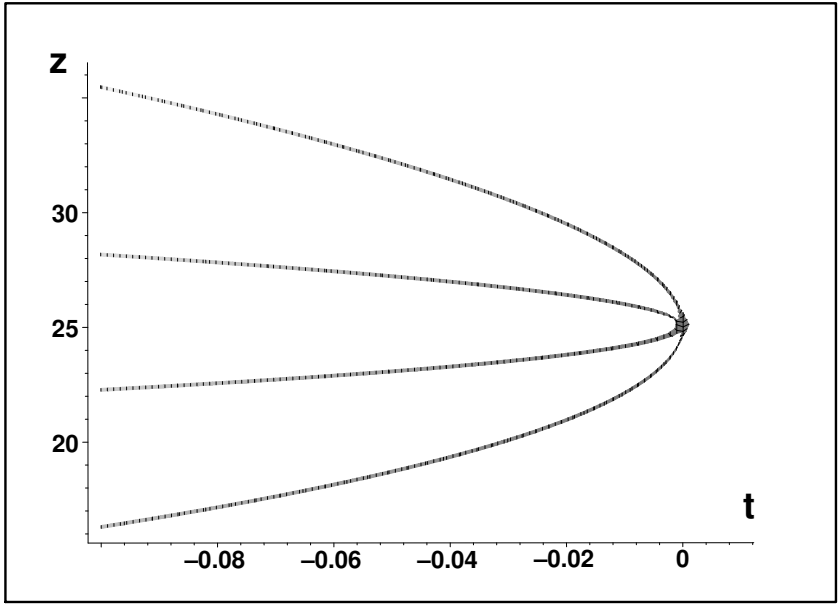}
\caption{The $t-$dependence of the real roots $z_j$ of secular
equation (\ref{seculae}). The collapse at $t=0$ is not destroyed due
to the weakness of the coupling to the environment ($\beta=1$).
 \label{obr0abc}}
\end{figure}

At the particular choice of $\beta=1$ the description of the
boundary $\partial {\cal D}_H$ remains feasible by non-numerical
means  yielding the transparent and algebraically tractable secular
equation
 \ben
 0=z^{4}+ \left(-100 -20\,t+2\,{t}^{2} \right) z^{3}+
 \een
 \be
 +\left(3750 +500\,t-80\,{t}^{2}-34\,{t}^{3} \right) z^{2}+
 \label{seculae}
  \ee
  $$+
 \left(-62500+ 12500\,t+4810\,{t}^{2}+360\,{t}^{3}+158\,{t}^{4}
 \right) z+
 $$
 $$
 +390625-312500\,t-23500\,{t}^{2}+22450\,
{t}^{3}-3221\,{t}^{4}-126\,{t}^{5}
 $$
which may very easily be treated numerically. Obviously, the level
$E_4=0$ separates while the other two quadruplets acquire the
square-root form $E_{4+j}=- E_{4-j}=\sqrt{z_j}$ for $j=1,2,3,4$.
Hence, one may proceed and study the spectrum of $z=E^2$ in full
parallel with our above $N=3$ model.


%
\begin{figure}
\includegraphics[width=8cm,angle=-90]{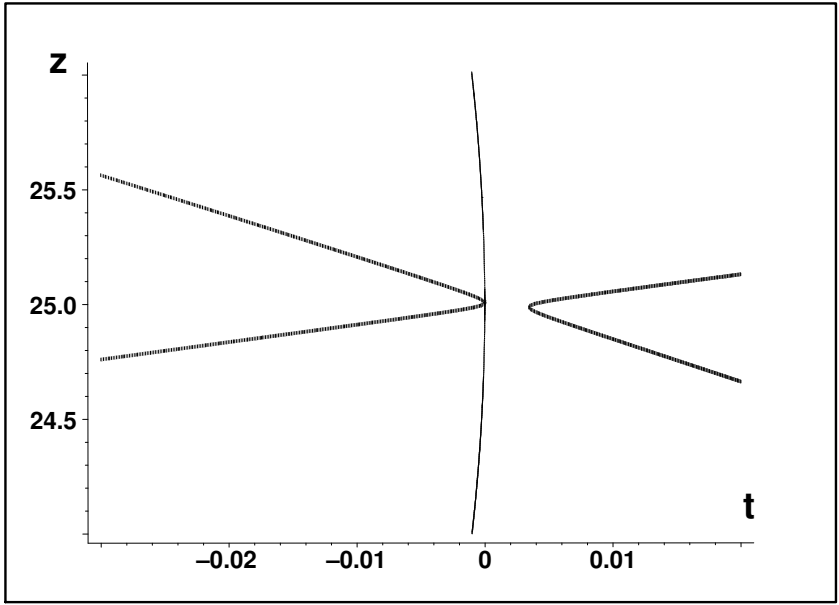}
\caption{The $t-$dependence of the real roots $z_j$ of secular
equation (\ref{seculae}) near $t=0$ at $\beta=2.73$. The collapse
survives, a partial recovery emerges at negative $t$.
 \label{obr0abcm}}
\end{figure}




The $\beta=1$ results are sampled in Fig.~\ref{obr0abc}. Inside the
physical domain of $t < 0$, qualitatively the same pattern is still
obtained even at the perceivably larger $\beta=2.73$ (cf.
Fig.~\ref{obr0abcm}). Once we are now getting very close to the
critical value of $\beta \approx 2.738$, the situation becomes
unstable. In the unphysical domain of $t>0$, for example, we can
spot an anomalous partial de-complexification of the energies at
certain positive values of parameter $t$.

%
%


%
\begin{figure}
\includegraphics[width=8cm,angle=-90]{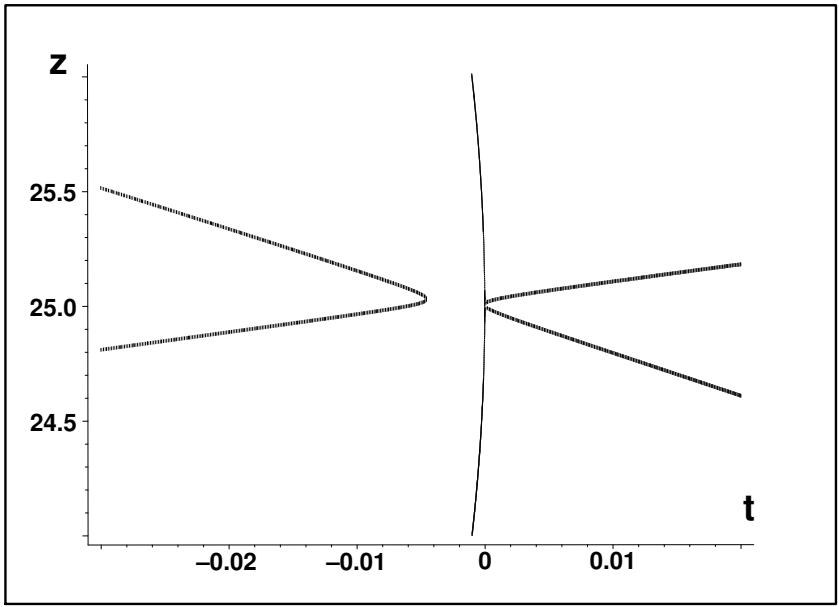}
\caption{The change of the $t-$dependence of the real roots $z_j$ of
secular equation (\ref{seculae}) near $t=0$ at $\beta=2.75$.
 \label{obr0abcv}}
\end{figure}

At cca $\beta \approx 2.738$ the two separate EP instants of the
degeneracy and complexification/decomplexification of the energies
fuse themselves. Subsequently, a qualitatively new pattern emerges.
Its graphical sample is given in Fig.~\ref{obr0abcv}. First of all,
the original multiple EP collapse gets decoupled. This implies that
at $\beta=2.75$ as used in the latter picture, the inner two levels
degenerate and complexify at a certain small but safely negative
$t=t_{crit}\approx -0.004$. Due to the solvability of the model we
may conclude that the boundary-curve $\partial {\cal D}_H$ starts
moving with  parameter $\beta$.

\section{Non-Hermitian quantum graphs\label{toymoN2}}

\subsection{Models with point interactions}

Another interesting ${\cal PT}-$symmetric single-particle
differential-operator Hamiltonian $H$ with the property $H \neq
H^\dagger$ in ${\cal H}^{(F)}$ has been proposed in
Ref.~\cite{David}. The particle of mass $\mu = 1/2$ has been assumed
there living on a finite interval of $x \in (-L,L)$. The only
nontrivial interaction was chosen as localized at the endpoints and
characterized by the Robin-type boundary conditions
 \begin{equation}
 \label{bc}
  \Psi'(\pm L) + i \alpha \;\! \Psi(\pm L) = 0\,,
  \ \ \ \ \alpha>0\,
   \ \ \ \ 2L\alpha/\pi
 \neq 1,2,\ldots
  \,.
 \end{equation}
The extreme simplicity of this model opened the way not only towards
the elementary formula for the energy spectrum,
 \begin{equation}
 \label{spectrum}
 E_0=\alpha^2\,,
 \ \ \
 E_n=\left (\frac{n\pi}{2L} \right )^2\,,
 \ \ \
 n=1,2,\ldots \,
 \end{equation}
but also  towards the equally elementary construction of the {\em
complete} family of the eligible metrics $\Theta$ (cf., e.g.,
Refs.~\cite{Jakub} for the details).

The solvability as well as extreme simplicity of this model proved
encouraging in several directions. In the present context, the
mainstream developments may be seen in the study of its discrete
descendants (cf. the next subsection). Nevertheless, before turning
our attention to the resulting family of the finite-dimensional
crypto-Hermitian problems, let us add a brief remark on the
alternative possibility of a transfer of the present analysis of the
idea of generalized solvability to the quickly developing field of
so called quantum graphs, i.e., of systems where the usual
underlying concept of a point particle moving along a real line or
interval is generalized in the sense that the single interval (say,
$e_+:=(0,L)$) is replaced by a suitable graph $\mathbb{G}^{(q)}$
composed of $q$ edges $e_j$, $j=0,1,\ldots,q-1$.

The idea still waits for its full understanding and consistent
implementation. In particular, in Ref.~\cite{canj} we showed that
even for the least complicated equilateral $q-$pointed star graphs
with $q>2$ the spectrum of energies need not remain real anymore,
even if one parallels, most closely, the $q=2$ boundary conditions
(\ref{bc}) and even if one does not attach any interaction to the
central vertex. In our present notation this means that the domain
${\cal D}_H$ of Fig.~\ref{obr0a} becomes empty. In other words, the
applicability of this and similar models remains restricted to
classical physics and optics while a correct, widely acceptable
quantum-system interpretation of the manifestly non-Hermitian $q>2$
quantum graphs must still be found in the future.

\subsection{Discrete lattices}

As we already indicated above, one of the most promising methods of
an efficient suppression of some of the above-mentioned shortcomings
of the ${\cal PT}-$symmetric models which are built in an
infinite-dimensional Hilbert space ${\cal H}^{(F)}$ may be seen in
the transition, say, to the discrete analogues and descendants of
various confining  ${\cal PT}-$symmetric as well as non$-{\cal
PT}-$symmetric potentials \cite{annals}. In particular, the most
elementary discrete analogues of the most elementary
end-point-interaction-simulating boundary conditions (\ref{bc}) may
be seen in the suitable end-point non-Hermitian perturbations
$W^{(N)}$ of the standard Hermitian kinetic-energy matrices
$-\triangle^{(N)}$, i.e., of the $N$ by $N$ negative discrete
Laplacean Hamiltonians where mere two diagonals of matrix elements
are non-vanishing,
$\triangle^{(N)}_{k,k+1}=\triangle^{(N)}_{k+1,k}=1$, $k = 1, 2,
\ldots, N-1$.

With this idea in mind we already studied, in Ref.~\cite{jmp}, the
most elementary model with
 \be
 {W}^{(N)}({\lambda})=\left[
 \begin {array}{cccccc}
     0&-{\it {\lambda}}&0&0&\ldots&0
 \\{}
 {\it {\lambda}}&0&0&0&\ddots&\vdots
 \\{}
 0&0&0&\ddots&\ddots&0
 \\{}
 0&0&\ddots&\ddots&0&0
 \\{}
 \vdots&\ddots&\ddots&0&0&{\it {\lambda}}
 \\{}
 0&\ldots&0&0&-{\it {\lambda}}&0
 \end {array}
 \right]\,.
 \label{ham11}
 \ee
We succeeded in constructing the complete $N-$parametric family of
the physics-determining solutions $\Theta$ of the compatibility
constraint (\ref{hiherua}). In Ref.~\cite{junde} we then extended
these results to the more general, multiparametric
boundary-condition-simulated perturbations
%
 \be
 {W}^{(N)}({\lambda, \mu})=\left[
 \begin {array}{cccccccc}
     0&-{\it {\lambda}}&0&0&\ldots&&\ldots&0
 \\{}
 {\it {\lambda}}&0&{\it {\mu}}&0&\ldots&&\ldots&0
 \\{}
 0&-{\it {\mu}}&0&0&\ddots&&&\vdots
 \\{}
 0&0&0&0&\ddots&\ddots&&\vdots
 \\{}
  \vdots&&\ddots&\ddots&\ddots&0&0&0
 \\{}
 \vdots &&&\ddots&0&0&-{\it {\mu}}&0
 \\{}
 0&\ldots&&\ldots&0&{\it {\mu}}&0&{\it {\lambda}}
 \\{}
 0&\ldots&&\ldots&0&0&-{\it {\lambda}}&0
 \end {array}
 \right]\,
 \label{ham11}
 \ee
etc. Thus, all of these models may be declared solvable in the
presently proposed sense. At the same time, the question of the
survival of feasibility of these exhaustive constructions of metrics
$\Theta$ after transition to nontrivial discrete quantum graphs
remains open~\cite{dgraphs}.

\section{Discussion\label{brokmo3}}

During  transitions from classical to quantum theory one must often
suppress various ambiguities -- cf., e.g., the well known
operator-ordering ambiguity of Hamiltonians which are, classically,
defined as functions of momentum  and position. Moreover, even after
we specify a unique quantum Hamiltonian operator $H$, we may still
encounter another, less known ambiguity which is well know, e.g., in
nuclear physics \cite{Geyer}. The mathematical essence of this
ambiguity lies in the freedom of our choice of a sophisticated
conjugation ${\cal T}^{(S)}$ which maps the standard physical vector
space ${\cal V}$ (i.e., the space of ket vectors $|\psi\kt$
representing the admissible quantum states) onto the dual vector
space ${\cal V}'$ of the linear functionals over ${\cal V}$. In our
present paper we discussed some of the less well known aspects of
this ambiguity in more detail. Let us now add a few further comments
on the current quantum-model building practice.

First of all, let us recollect that one often postulates a
point-particle (or point-quasi-particle) nature and background of
the generic quantum models. Thus, in spite of the existence of at
least nine alternative formulations of the abstract quantum
mechanics as listed, by Styer et al, in their 2002 concise review
paper \cite{amj}, a hidden reference to the wave function $\psi(x)$
which defines the probability density and which lives in some
``friendly'' Hilbert space (say, in ${\cal
H}^{(F)}=L^2(\mathbb{R}^d)$) survives, more or less explicitly, in
the large majority of our conceptual as well as methodical
considerations.

A true paradox is that the  {\em simultaneous} choice of the
friendly Hilbert space ${\cal H}^{(F)}$ {\em and} of some equally
friendly differential-operator generator $H = \triangle+V(x)$ of the
time evolution encountered just a very rare critical opposition in
the literature \cite{clock}. The overall paradigm only started
changing when the nuclear physicists imagined that the costs of
keeping the Hilbert space ${\cal H}^{(F)}$ (or, more explicitly, its
inner product) unchanged may prove too high, say, during variational
calculations \cite{Geyer}. Anyhow, the ultimate collapse of the old
paradigm came shortly after the publication of the Bender's and
Boettcher's letter \cite{BB} in which, for  certain friendly ODE
Hamiltonians  $H = \triangle+V(x)$ the traditional choice of {space}
${\cal H}^{(F)}=L^2(\mathbb{R})$ has been found {unnecessarily}
over-restrictive (the whole story may be found described in
\cite{Carl}).

The net result of the new developments may be summarized as an
acceptability of a less restricted input dynamical information about
the system. In other words, the use of the friendly space ${\cal
H}^{(F)}$ {\em in combination with} a friendly Hamiltonian
$H=H^\dagger$ has been found a theoretician's luxury. The need of a
less restrictive class of standard Hilbert spaces ${\cal H}^{(S)}$
which would differ from their ``false'' predecessor ${\cal H}^{(F)}$
by a nontrivial inner-product metric $\Theta\neq I$ appeared
necessary.

One need not even abandon the most common {\em a priori} selection
of the friendly Hilbert space ${\cal H}^{(F)}$ of the ket vectors
$|\psi\kt$ with their special Dirac's duals (i.e., roughly speaking,
with the transposed and complex conjugate bra vectors $\br \psi|$)
yielding the Dirac's inner product $\br \psi_1|\psi_2\kt=\br
\psi_1|\psi_2\kt^{(F)}$. What is only new is that such a
pre-selected, $F-$superscripted Hilbert space need not necessarily
retain the usual probabilistic interpretation.

One acquires an enhanced freedom of working with a sufficiently
friendly form of the input Hamiltonian $H$, checking solely the
reality of its spectrum. Thus, one is allowed to admit that $H \neq
H^\dagger$ in ${\cal H}^{(F)}$. One must only introduce, {\em on
some independent initial heuristic grounds}, the amended Hilbert
space ${\cal H}^{(S)}$. For such a purpose it is sufficient to keep
the same ket-vector space and just to endow it with some
sufficiently general and Hamiltonian-adapted (i.e.,
Hamiltonian-Hermitizing) inner product (\ref{ip}) \cite{Geyer}. This
is the very core of innovation. In the physical Hilbert space ${\cal
H}^{(S)}$  the unitarity of the evolution of the system must remain
guaranteed, as usual, by the Hermiticity of our Hamiltonian {\em in
this space}, i.e., by a hidden Hermiticity condition
 \be
 H=\Theta^{-1}H^\dagger\Theta:=H^\ddagger\,
 \label{dieudd}
 \ee
{\it alias} crypto-Hermititicity condition \cite{SIGMA}. In the
special case of finite matrices one speaks about the
quasi-Hermiticity condition. Unfortunately, the latter name becomes
ambiguous and potentially misleading whenever one starts
contemplating certain sufficiently wild operators in general Hilbert
spaces~\cite{Dieudonne}.

It is rarely emphasized (as we did in \cite{fortschr}) that the
choice of the metric remains an inseparable part of our
model-building duty {\em even if } our Hamiltonian happens to be
Hermitian, incidentally, {\em also} in the unphysical initial
Hilbert space ${\cal H}^{(F)}$.  {\em Irrespectively} of the
Hermiticity or non-Hermiticity of $H$ in auxiliary ${\cal H}^{(F)}$,
one {\em must} address the problem of the {independence} of the
dynamical input information carried by the metric $\Theta$. Only the
simultaneous specification of the operator pair of $H$ and $\Theta$
connected by constraint (\ref{dieudd}) defines physical predictions
in consistent manner. In this sense, the concept of solvability must
necessarily involve also the simplicity of $\Theta$.

\subsection*{Acknowledgements}

Work supported by GA\v{C}R, grant Nr. P203/11/1433.

\end{document}